\documentclass[aps,pre,reprint,showpacs,floatfix,superscriptaddress]{revtex4-1}
\usepackage[colorlinks,linkcolor=red,anchorcolor=red,citecolor=blue,urlcolor=blue]{hyperref}
\usepackage{amsmath,txfonts,sidecap}
\usepackage{graphicx}% Include figure files
\usepackage{dcolumn,enumitem}% Align table columns on decimal point
\usepackage{float}
\allowdisplaybreaks[3]

\begin{document}

\title{Temporal link prediction methods based on behavioral synchrony}

\author{Yueran Duan}
\email{sarah\_623@outlook.com}
\affiliation{School of Economics and Management, China University of Geosciences, Beijing, China}
\affiliation{Department of Computer Science, Aalto University, Finland}
\author{Qing Guan}
\email{guanqing35@126.com}
\affiliation{School of Information Engineering, China University of Geosciences, Beijing, China}
\author{Petter Holme}
\email{petter.holme@aalto.fi}
\affiliation{Department of Computer Science, Aalto University, Finland}
\author{Yacheng Yang}
\email{yang-yc21@mails.tsinghua.edu.cn}
\affiliation{School of Economics and Management, Tsinghua University, China}
\author{Wei Guan}
\email{guanwei1996@outlook.com}
\affiliation{School of Economics and Management, China University of Geosciences, Beijing, China}

% printed abstract
\begin{abstract}
Link prediction---to identify potential missing or spurious links in temporal network data---has typically been based on local structures, ignoring long-term temporal effects. In this chapter, we propose link-prediction methods based on agents' behavioral synchrony. Since synchronous behavior signals similarity and similar agents are known to have a tendency to connect in the future, behavioral synchrony could function as a precursor of contacts and, thus, as a basis for link prediction. We use four data sets of different sizes to test the algorithm's accuracy. We compare the results with traditional link prediction models involving both static and temporal networks. Among our findings, we note that the proposed algorithm is superior to conventional methods, with the average accuracy improved by approximately 2\%-5\%. We identify different evolution patterns of four network topologies---a proximity network, a communication network, transportation data, and a collaboration network. We found that: (1) timescale similarity contributes more to the evolution of the human contact network and the human communication network; (2) such contribution is not observed through a transportation network whose evolution pattern is more dependent on network structure than on the behavior of regional agents; (3) both timescale similarity and local structural similarity contribute to the collaboration network.
\end{abstract}

\maketitle

\section{Introduction}
\label{sec:introduction}

Homophily is an important mechanism for network link-formation in systems as diverse as friendships, collaborations, investments, information diffusion, citations, etc.~\cite{garrod2021influencing,mcpherson2001birds}. Furthermore, some studies argue that synchronous behavior between agents indicates a similarity~\cite{dong2015actors,wiltermuth2009synchrony}. For example, the times for stock-market agents (investors, fund managers, companies, etc.)\ to respond to financial distress, policy change, etc., could divide them into groups of different abilities---agents able to respond similarly show homophily throughout a data set~\cite{chakrabarti2021covid}. This phenomenon could also happen in trade~\cite{meshcheryakova2020similarity}, collaboration~\cite{li2020evolution}, information diffusion, communication~\cite{riad2019risk}, and other scenarios.

For a decade, link prediction has been an active area of research~\cite{lu2011link}. One common principle is to exploit neighborhood-based similarity. There are two commonly used methods: First, first-order neighborhood methods, such as common neighbor index~\cite{liben2003link} and preferential attachment index~\cite{barabasi1999emergence}. Second, second-order neighborhood methods, such as the Adamic-Adar index~\cite{adamic2003friends} and resource allocation index~\cite{zhou2009predicting}.

Several studies have incorporated temporal information into link prediction models to further test the time effects on network evolution~\cite{ahmed2009recovering}. Recent progress in temporal link prediction involves four methods: time decay function~\cite{liao2017ranking,butun2018extension}, matrix factorization-based prediction model~\cite{ma2017nonnegative,wu2018tracking}, deep learning~\cite{xiang2020ti}, and probability model~\cite{ouzienko2010prediction}. The time decay function is popular in empirical research due to its relatively light computational complexity, but two issues remain. First, the behavior of an agent has the same effect on the link prediction, irrespective of time. Second, the time decay function cannot capture the temporal changes in the similarity between agents.

This research provides what we call a \textit{local similarity time vector} (LSTV) link prediction method, which quantitatively captures the behavioral synchrony of agents and utilizes such information to design an effective and universal link prediction method for temporal networks. We depicted the effects of time on network evolution by providing a timescale similarity index that measures the similar contacts at points in time between two agents. Then, we add the timescale similarity index to the link prediction model to predict the potential linkage. We selected four data sets of different sizes to test the accuracy of our algorithm. We also compared the results with static network link prediction models and competing typical temporal link prediction models. We also distinguished topologies of networks with different evolution patterns by exploring the combination of two similarity measures. This research contributes to the effects of agents' behavior synchrony on network evolution by adding the time information into the network and improving the prediction accuracy of the LSTV model using the temporal similarity index.

The rest of this chapter is arranged as follows: Section \ref{sec: problem} introduces the problem statement, and its evaluation metrics; Section \ref{sec: related} reviews some related works; Section \ref{sec: from} presents the LSTV link prediction model; Section \ref{sec:data} presents the data sets; Section \ref{sec:expe} designs the experiments and analyzes the results; Section \ref{sec: conclusion} provides the conclusions.

\section{Problem statement and evaluation metrics}
\label{sec: problem}

\subsection{Temporal Link Prediction}
\label{sec: temporal link prediction}
Consider an undirected temporal network $G(V,E,T)$, where $V=\{v_1, v_2,\dots, v_n\}$, $E=\{e_1, e_2,\dots, e_n\}$, and $T=\{\tau_1, \tau_2, \dots, \tau_n\}$ denote the sets of nodes, links, and time stamps, respectively. Additionally we use the notations, $n=|V|$, $m=|E|$, $k=|T|$. We do not consider self-connections and multiple links. 

$A_{\tau_h}$ is an adjacency matrix of $G_{\tau_h}$:
\begin{equation} \label{eq:adjacency matirx}
A_{\tau_h} = \begin{bmatrix}
  a_{11\tau_h} & \cdots & a_{1n\tau_h} \\
  \vdots & \ddots  & \vdots \\
  a_{n1\tau_h} &  \cdots  & a_{nn\tau_h}
\end{bmatrix}
\end{equation}
where $a_{ij\tau_h}=1$ if there is a link between $v_i$  and $v_j$  at time $\tau_h$ , if not, $a_{ij}=0$ .

Correspondingly, each node has a time vector according to the time of node interaction. The temporal activity of node $v_i$  can be represented by:
\begin{equation} \label{eq:temporal activity}
T_i=(d_{i1},d_{i2},\dots,d_{ik})
\end{equation}
where, $d_{ij}$  indicates the degree of node $v_i$  at time $\tau_j$ .

According to this definition, we attempt to forecast the potential link at time $\tau_{k+1}$. Definitions are listed in Table~\ref{table:variables} for reference.

\subsection{Evaluation Metrics}
\label{sec: evaluation metrics}
We choose AUC and ranking score as the performance evaluation indexes of the algorithm. In general, we delete 10\% edges randomly from the networks and the remaining 90\% edges make up the training set $E^T$.

AUC is defined as the probability that the randomly selected missing link is given a higher score than a randomly chosen nonexistent link~\cite{lu2011link}. After independent comparison for n times, AUC value can be defined as:
\begin{equation} \label{eq:AUC}
AUC=\frac{n^\prime+0.5n^{\prime \prime}}{n}
\end{equation}
where $n^\prime$  indicates the number of times that the score of the link in $E^P$  is greater than that of the edge in $E^N$. $n^{\prime \prime}$  indicates the number of times that the score of the link in $E^P$  is equal to that of the link in $E^N$. According to Eq.~\ref{eq:AUC}, a higher value of AUC indicates the degree to which the potential linking mechanism of the algorithm is more accurate than the random selection. $AUC=0.5$  indicates the score is random.

Ranking score measures the ranking of links in the probe set of unknown links. Let $S=U-E^T$, $S$ is the set of unknown links, including both in the probe set and non-existent links. The \textit{ranking score} is defined as~\cite{zhou2007bipartite}:
\begin{equation} \label{eq:RankS}
{\rm RankS} = \frac{{\textstyle \sum_{i\in E^P}^{} {\rm RankS}_i} }{| E^P |} =  \frac{1}{| E^P |}  {\textstyle \sum_{i\in E^P}\frac{r_i}{| S |}} 
\end{equation}
where $r_i$  represents the rank of the edge $i$'s score in the set of unknown edges. According to this equation, a smaller value RankS represents a higher edge ranking in the probe set, which means that the probability of being predicted successfully is higher so the algorithm's accuracy is higher.

\begin{table*}[htbp]
	\centering
	\caption{\centering \label{table:variables} Variables used to deal with each data unit}
	\begin{tabular}{cp{4.5cm}cp{4.5cm}} %居中：p{5cm}<{\centering}
		\toprule  % 顶部线
		Variables & Definition & Variables & Definition \\ 
		\colrule  % 中部线
		$V$ & the set of nodes & $n$ & the number of nodes\\
            $E$ & the set of edges & $m$ & the number of edges\\
            $T$ & the set of time stamps & $k$ & the number of time stamps \\
            $E^T$ & the set of training edges & $E^p$ & the set of probe edges \\
            $E^N$ & the set of nonexistent edges & & \\
            $U$ & the universal link set of all nodes ($U=\frac{N\times (N-1)}{2}$ for undirected network)  & $s_{xy}$ & the link prediction score of the link between node $v_x$  and node $v_y$ \\
            $\Gamma(x)$ & the neighbor set of node $v_x$ & $k_x$ & the degree of node $v_x$\\
		\botrule  % 底部线
	\end{tabular}
\end{table*}

\section{Related work}\label{sec: related}

\subsection{Link prediction in static network}
\label{sec: link prediction in static network}
The purpose of link prediction is to detect missing links and identify spurious interactions from existing parts of the network, and we can try to reconstruct the observed networks by using these models~\cite{guimera2009missing}. A number of prediction models have been developed, but the simplest framework for link prediction is based on the similarity between nodes for design and prediction~\cite{lu2011link}. The higher the similarity between two nodes is, the higher the probability that two nodes may have a link~\cite{newman2001clustering}. For example, nodes with more common neighbors tend to link probably~\cite{liben2003link}. In the field of similarity-based link prediction, there are three main methods: local index, quasi-local index, and global index~\cite{lu2011link}.

Through the verification of a large number of empirical networks, scholars found that the link prediction method based on local information similarity has a better performance, which is also simple and easy to operate~\cite{lu2011link}. Table~\ref{table:indicators_LS} lists five widely used link prediction models based on local structural similarity. The five methods will be used as a comparative test to verify the superiority of the algorithm proposed in this chapter.
\begin{table*}[htbp]
	\centering
	\caption{\centering \label{table:indicators_LS} Indicators used to describe local similarity (LS)}
	\begin{tabular}{p{2.5cm}p{3.5cm}p{5.5cm}} %居中：p{5cm}<{\centering}
		\toprule  % 顶部线
		Local similarity & \quad Function & Description \\ 
		\colrule  % 中部线
		Common Neighbor Index (CN) &  \quad $s_{xy}^{CN}=| \Gamma (x) \cap \Gamma (y) |$  & CN takes the number of common neighbors of two nodes as the similarity score~\cite{liben2003link}. The calculation of this indicator is simple, and the performance is competitive.\\
        Jaccard Coefficient (JC) & \quad $s_{xy}^{JC}=\frac{| \Gamma (x) \cap \Gamma (y) |}{| \Gamma (x) \cup \Gamma (y) |}$ &  JC uses the ratio of the common neighbor of two nodes to all the neighbors of the two nodes as the similarity score~\cite{jaccard1901etude}.\\
        Preferential Attachment Index (PA) & \quad $s_{xy}^{PA}=k_x k_y$ & PA reflects the principle of preferential attachment~\cite{barabasi1999emergence}.\\
        Adamic-Adar Index (AA) & \quad $s_{xy}^{AA} = \sum_{z\in | \Gamma (x) \cap \Gamma (y) | }^{}\frac{1}{\log k_z}$ & AA index considers that vertices with fewer shared relations have higher linking probability~\cite{adamic2003friends}.\\
        Resource Allocation (RA) & \quad $s_{xy}^{AA} = \sum_{z\in | \Gamma (x) \cap \Gamma (y) | }^{}\frac{1}{k_z}$ & RA is similar to AA in its way of empowerment~\cite{zhou2009predicting}.\\
		\botrule  % 底部线
	\end{tabular}
\end{table*}

\subsection{Link prediction in temporal networks}
\label{sec: link prediction in temporal network}
Link prediction in temporal networks is a method of evaluating network evolving mechanisms by obtaining network data that evolves over time, and then finding future links. Due to the time-varying characteristics of complex systems, more scholars tried incorporating time information into the link prediction model. At present, there are several temporal link prediction methods: deep learning, matrix factorization-based prediction model, time decay function, etc.

\subsubsection{Temporal link prediction with deep learning}

Graph embedding can retain the properties of temporal networks while mapping the networks into low-dimensional vector spaces~\cite{xiang2020ti}. Network embedding can solve the problem that graph data is difficult to input into machine learning algorithms efficiently.

With the development of graph neural networks, deep learning frameworks such as graph convolutional network (GCN), long short-term memory network (LSTM), and generative adversarial networks (GAN) are gradually applied to the dynamic prediction of complex networks~\cite{lei2019gcn}. In particular, Lei et al.~\cite{lei2019gcn} used GCN to explore the local topological features of each time segment, used LSTM model to represent the evolution characteristics of dynamic networks, and finally used GAN framework to enhance the model to generate the next weighted network segment. Based on this model, many improved algorithms have attracted more scholars' attention, such as adding time decay coefficient~\cite{meng2019nelstm}, increasing attention mechanism~\cite{yang2019advanced}, and so on.

Graph embedding is more practical than adjacency matrices because they pack node attributes into a vector with smaller dimensions. Vector operations are simpler and faster than graphically comparable operations. However, the prediction model based on deep learning is complicated in design, its potential mechanism is unclear, and it is not universal in different types of networks. Although these models perform well, they are not widely used in empirical studies.

\subsubsection{Matrix factorization-based temporal link prediction}

Link prediction can be regarded as an adjacency matrix filling problem, which can be solved by combining explicit and implicit features of nodes and links with matrix decomposition method by bilinear regression model~\cite{menon2011link}. Because matrix/tensor is the most direct method to characterize networks, matrix/tensor decomposition-based link prediction model is widely used in temporal link prediction. Ma et al.~\cite{ma2017nonnegative} proved the equivalence between the nonnegative matrix decomposition (NMF) of a communicability matrix and the eigendecomposition (ED) of the Katz matrix, providing a theoretical explanation for the application of NMF to the temporal link prediction problem.

Dunlavy et al.~\cite{dunlavy2011temporal} expressed the link prediction problem as a periodic temporal link prediction, compressed the data into a single matrix through weight assignment, and obtained the time characteristics by tensor decomposition. Ma et al.~\cite{ma2018graph} used graph communicability to decompose each network to obtain features and then collapsed the feature matrix to predict temporal links. However, these studies all decomposed network or collapsed features at each time layer and ignore the relationship among slices. To overcome this problem, scholars proposed a graph regularization non-negative matrix decomposition algorithm for temporal link prediction without destroying dynamic networks~\cite{ma2018graph,zhang2019graph}. Based on the graph regularization non-negative matrix decomposition algorithm, some scholars fused the community topology as a piece of prior information into a new temporal link prediction model, effectively improving the prediction accuracy~\cite{zhang2021semi}.

Compared with the prediction model based on deep learning methods, the matrix decomposition-based prediction model has much fewer parameters to learn. However, it is difficult to be used in large-scale networks and empirical studies due to the high cost of matrix/tensor calculation.

\subsubsection{Link prediction with a time decay function}

During the growth of the network, agents exhibit heterogeneous fitness values that decay over time~\cite{medo2011temporal}. Therefore, many scholars have incorporated the time decay function into the link prediction model based on structural similarity to obtain a better prediction effect~\cite{liao2017ranking,butun2018extension}. Meanwhile, some scholars added the time decay function as the weight to the node pair for graph embedding and finally added it into the framework of deep learning~\cite{meng2019nelstm}.

The time decay effect was proposed because scholars focused on the mechanism that some effects weaken over time during the evolution of networks. The time decay function-based prediction model is simple and performs well in some networks. However, the time decay function appears as the weight of some attributes of networks and does not distinguish the network structure of different time layers. The time-based link prediction method proposed in this chapter also adds the time factor to the original prediction model, so the prediction model with the time decay factor is mainly selected for comparison experiments.

\subsubsection{Other temporal link prediction models}

In addition to the above models, some temporal link prediction models, such as time series-based prediction models, probabilistic prediction models, and higher-order structure prediction models, have also been well developed.

The time series-based temporal link prediction model generally builds time series based on various node similarity measures and scores the possibility of node pair linking by time series prediction model~\cite{da2012time,huang2009time,yang2015time}. One case is that G\"unecs et al.~\cite{gunecs2016link} combined some Autoregressive (AR) and Moving Average (MA) processes as time series prediction models. Probabilistic models quantify change and uncertainty by using maximum likelihood methods or probability distributions. The probability distribution describes the range of possible values and gives the most likely values, which is helpful in considering all possible outcomes in the process of temporal link prediction. Such methods focus on constructing probabilistic models, such as the Exponential Random Graph Model (ERGM)~\cite{ouzienko2010prediction}. To better simulate real interaction scenarios, some scholars have developed temporal link prediction methods based on motif features or higher-order features to help us understand and predict the evolution mechanism of different systems in the real world~\cite{yao2021higher,8489644,eun_tnwk}. 

\section{From time decay function to time vector}
\label{sec: from}

\subsection{Local similarities with a temporal logarithmic decay function (LSTD) link prediction model}
\label{sec: LSTD model}
The LSTD link prediction model uses a time decay function as the weight of the local structure similarity score. The model is simple and common in design and can be embedded into some complex models. Therefore, the LSTD prediction model is popular in both empirical research and algorithm improvement.

This chapter chooses the LSTD link prediction model with a temporal logarithmic decay function as a comparison experiment of link prediction algorithms. The temporal function can be defined as~\cite{butun2018extension}:
\begin{equation} \label{eq:temporal function}
F_t(v_x,v_z)=\log[(t(v_x,v_z)-t_{start})+c]
\end{equation}
where $c$  is a constant. Then, the time function is added as a weight to the similarity indicators based on common neighbors. The formulas are shown in Table~\ref{table:LSTD}:
\begin{table*}[htbp]
	\centering
	\caption{\centering \label{table:LSTD} Functions used to describe LSTD link prediction model}
	\begin{tabular}{cp{8cm}} %居中：p{5cm}<{\centering}
		\toprule  % 顶部线
		Algorithm & \quad Temporal extensions \\ 
		\colrule  % 中部线
		CN &  \quad $s_{xy}^{TCN}= { \sum_{v_z\in \Gamma (x) \cap \Gamma (y)}^{}f_t(v_x,v_z)+f_t(v_y,v_z)}$ \\
            JC & \quad $s_{xy}^{TJC}= { \sum_{v_z\in \Gamma(x) \cap \Gamma(y)}^{}\frac{f_t(v_x,v_z)+f_t(v_y,v_z)}{ { \sum_{v_a\in \Gamma(x)}^{}f_t(v_x,v_a)}+ { \sum_{v_b\in \Gamma(y)}^{}f_t(v_y,v_b)}}}$\\
            PA & \quad $s_{xy}^{TPA}= { \sum_{v_a\in \Gamma(x)}^{}f_t(v_x,v_a)} \times  { \sum_{v_b\in \Gamma(y)}^{}f_t(v_y,v_b)}$\\
            AA & \quad $s_{xy}^{TAA}= { \sum_{v_z\in \Gamma(x)\cap \Gamma(y)}^{}\frac{f_t(v_x,v_z)+f_t(v_y,v_z)}{\log(1+ { \sum_{v_c\in \Gamma(z)}^{}f_t(v_z,v_c)} )}}$\\
            RA & \quad $s_{xy}^{TRA}= { \sum_{v_z\in \Gamma(x)\cap \Gamma(y)}^{}\frac{f_t(v_x,v_z)+f_t(v_y,v_z)}{\mathrm{} { \sum_{v_c\in \Gamma(z)}^{}f_t(v_z,v_c)})}}$\\
		\botrule  % 底部线
	\end{tabular}
\end{table*}

\subsection{Local similarities and temporal vector (LSTV) link prediction model}
\label{sec: LSTV model}
This chapter proposes a local similarity with a temporal vector (LSTV) link prediction algorithm. The temporal vector of each agent is generated according to whether the agent interacts at each point of time, and the similarity measure of a temporal vector is added to the original link prediction model. 

\begin{figure}
  \includegraphics[width=\columnwidth]{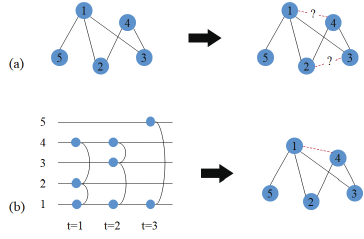}  
\caption{A sample of the local similarity with temporal vector link prediction algorithm.
(a) Calculate the common neighbors of unknown links: $s_CN(1,4)=2$; $s_CN(2,3)=2$; $s_CN(2,5)=1$; $s_CN(3,5)=1$; $s_CN(4,5)=0$. The most potentially possible links are 1-4 and 2-3 ($s_CN(1,4)=s_CN(2,3)=2$).
(b) Calculate the temporal vectors of each node: $1:[1,1,1]$, $2:[2,0,0]$, $3:[0,2,0]$, $4:[1,1,0]$, $5:[0,0,1]$. The timescale similarity between nodes is: $\cos(1,4)=0.667$; $\cos(2,3)=0$. Because $\cos(1,4)>\cos(2,3)$, then the most potentially possible link is 1-4.}
\label{fig:LSTV}
\end{figure}

Specifically speaking, in the link prediction process based on common neighbors, the common neighbors of node pairs are positive integers. Many node pairs have the same number of common neighbors whose priority is indistinguishable. In this case, the degree of similarity between agents can be judged by analyzing their behavior synchrony at different times. As shown in Fig.~\ref{fig:LSTV}, the network generation has $n$ time periods, so each agent has an $n$-dimensional temporal vector. The temporal vector of each agent is obtained by marking $d$ when interaction occurs $d$ times at time 1, and 0 when interaction does not occur at the next time. Then, we choose the cosine of the temporal vectors of the two agents as the measurement of the similarity between the agents and add it to the traditional link prediction model based on the neighbor similarity.

Timescale similarity of behaviors between agents can be defined as:
\begin{equation} \label{eq:timescale similarity}
t_{xy}=\cos(T_x,T_y)=\frac{ { \sum_{i=1}^{k}(t_{xi}\times t_{yi})} }{\sqrt{ { \sum_{i=1}^{k}(t_{xi})^2}}\times \sqrt{ { \sum_{i=1}^{k}(t_{yi})^2}}   } 
\end{equation}

The link prediction score of this method can be defined as:
\begin{equation} \label{eq:prediction score}
{\rm Sim}_{xy}^{LSTV}=\alpha \times \frac{t_{xy}}{t_{\max}} +(1-\alpha)\times \frac{s_{xy}}{s_{\max}}, \alpha \in [0,1] 
\end{equation}
where $t_{\max}$  represents the maximum of the timescale similarity of all node pairs, and $s_{xy}$ represents the link prediction score based on a certain local structure. Accordingly, $s_{\max}$  represents the maximum score of all node pairs under the link prediction model.

\subsection{Local similarities with a temporal logarithmic decay function and temporal vector (LSTDV) link prediction model}
\label{sec: LSTDV model}
The link prediction method based on the time decay function assumes that the homophily effect of the agent will decrease with time. LSTD link prediction model takes the time decay function as the weight of local structure similarity between agents. The agents' timescale similarity proposed in this chapter can also be used as an index of the similarity between agents. Then, does timescale similarity have a decay effect? Therefore, based on the LSTD link prediction model, we also take the time decay function as the weight of the timescale similarity index to explore the decay effect of agent behavior synchrony.

The temporal activity of node $v_i$  after adding the time decay function can be expressed as:
\begin{equation} \label{eq:temporal activity_LSTD}
T_i^D=(\log(1+c)d_{i1}, \log(2+c)d_{i2},\dots ,\log(k+c)d_{ik})
\end{equation}

Timescale similarity with the time decay function of behaviors between agents can be defined as:
\begin{equation} \label{eq:timescale similarity_LSTD}
t_{xy}^D=\cos(T_x^D,T_y^D)
\end{equation}
The link prediction score of this method can be defined as:
\begin{equation} \label{eq:prediction score_LSTD}
{\rm Sim}_{xy}^{LSTDV}=\alpha \times \frac{t_{xy}^D}{t_{\max}^D} +(1-\alpha)\times \frac{s_{xy}^D}{s_{\max}^D}, \alpha \in [0,1] 
\end{equation}
where $t_{\max}^D$  represents the maximum of the timescale similarity with the time decay function of all node pairs.

\subsection{Local similarities temporal vector for heterogeneous time layer (LSTHV) link prediction model}
\label{sec: LSTHV model}
The LSTDV model distinguishes the importance of time layers in the form of time decay. The AA index considers that the contribution of less-connected common neighbors is greater than that of more-connected common neighbors~\cite{adamic2003friends}. Inspired by the AA index, we believe that the less-connected time layer contributes more than the more-connected time layer. Therefore, we proposed the LSTHV prediction model, and the importance of the time layer can be expressed as:
\begin{equation}\label{time layer importance}
    h_i=\frac{1}{\log(m_i+c)}
\end{equation}
where $h_i$  represents the importance of time layer $i$, $m_i$ represents the number of time layer $i$'s edges, and $c$ is a constant. The temporal activity of node $v_i$  after adding the importance of the time layer can be expressed as:
\begin{equation}\label{temporal activity_LSTHV}
    T_i^H=(h_1d_{i1}, h_2d_{i2},\dots ,h_kd_{ik})
\end{equation}

Timescale similarity for heterogeneous time layers can be defined as:
\begin{equation}\label{timescale similarity_LSTHV}
    t_{xy}^H=\cos(T_x^H,T_y^H)
\end{equation}

The link prediction score of this method can be defined as:
\begin{equation}\label{prediction score_LSTHV}
    {\rm Sim}_{xy}^{LSTHV}=\alpha \times \frac{t_{xy}^H}{t_{\max}^H} +(1-\alpha)\times \frac{s_{xy}^H}{s_{\max}^H}, \alpha \in [0,1]
\end{equation}
where $t_{\max}^H$  represents the maximum of the timescale similarity with the importance of the time layer of all node pairs.

\section{Data}
\label{sec:data}
In this chapter, four different temporal networks are selected to verify the algorithm. They are 1) Hypertext; 2) UK airline; 3) Hollywood; 4) EU email. These four networks have been briefly introduced in the text above.

\begin{figure}
  \includegraphics[width=\columnwidth]{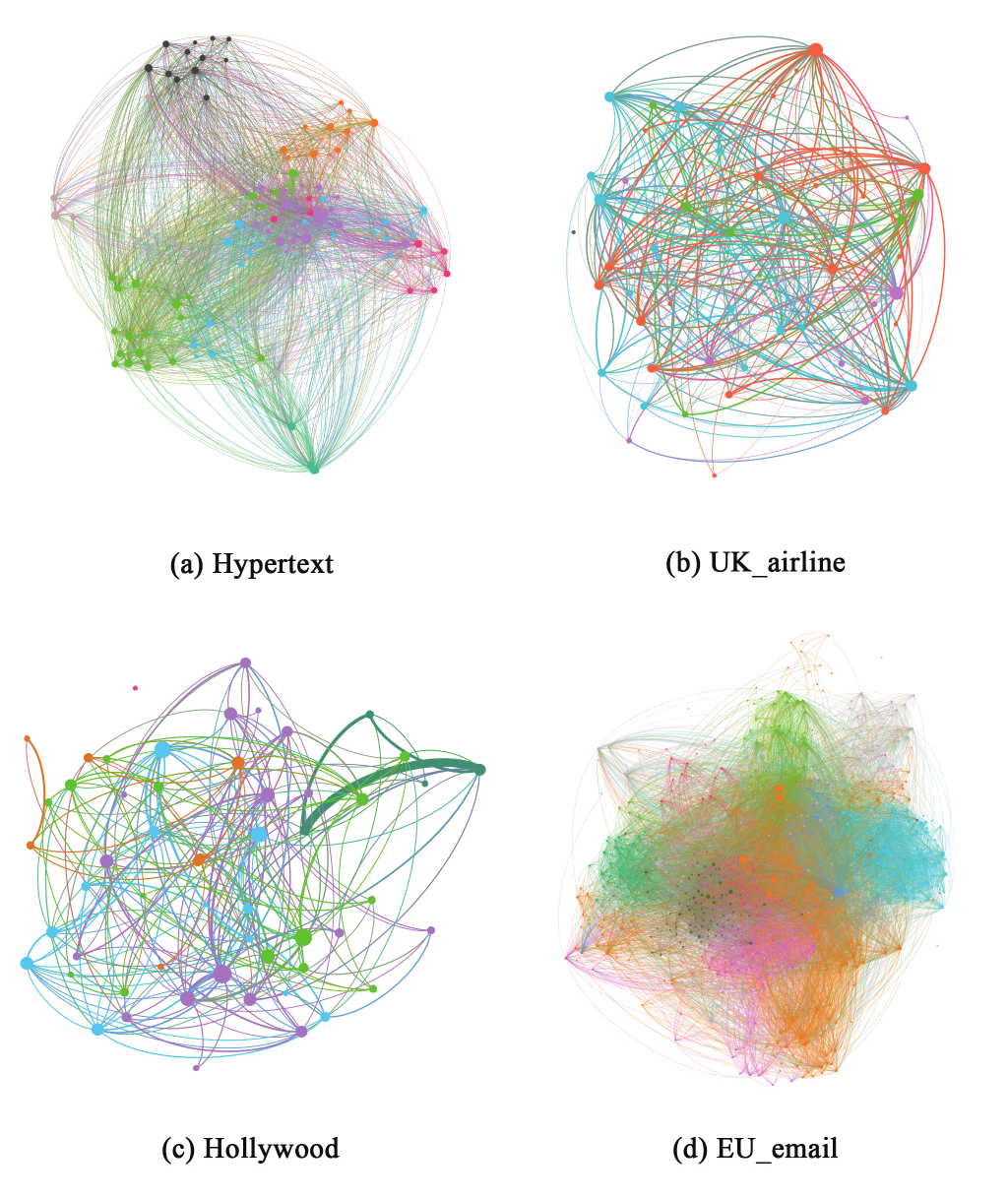}  
\caption{Visualization of the four empirical networks. Colors represent different communities, obtained by a network clustering algorithm (for the purpose of this paper, they only serve as decoration).}
\label{fig:visualization}
\end{figure}

We calculated the average degree (AD), average weighted degree (AWD), network diameter (ND), density, average clustering coefficient (ACC), and average path length (APL) of the four networks in static state, as shown in Table~\ref{table:global characteristics}. The four network visualization pictures are shown in Fig.~\ref{fig:visualization}. Among them, Hypertext and Eu Email have obvious community structures, while UK Airline shows typical geographical characteristics.
\begin{table}[htbp]
	\centering
	\caption{\centering \label{table:global characteristics} Summary statistics of the four data sets}
	\begin{tabular}{ccccccc} %居中：p{5cm}<{\centering}
		\toprule  % 顶部线
		Dataset & AD & AWD & ND & Density & ACC & APL\\ 
		\colrule  % 中部线
		Hypertext & 38.867 & 368.46 & 3 & 0.347 & 0.54 & 1.656\\
            UK airline & 14.764 & 109.673 & 4 & 0.273 & 0.699 & 1.833\\
            Hollywood & 8.4 & 21.345 & 4 & 0.156 & 0.331 & 2.161\\
            EU email & 30.357 & 638.142 & 7 & 0.031 & 0.439 & 2.622\\
		\botrule  % 底部线
	\end{tabular}
\end{table}

\begin{figure*}
  \includegraphics[width=0.8\textwidth]{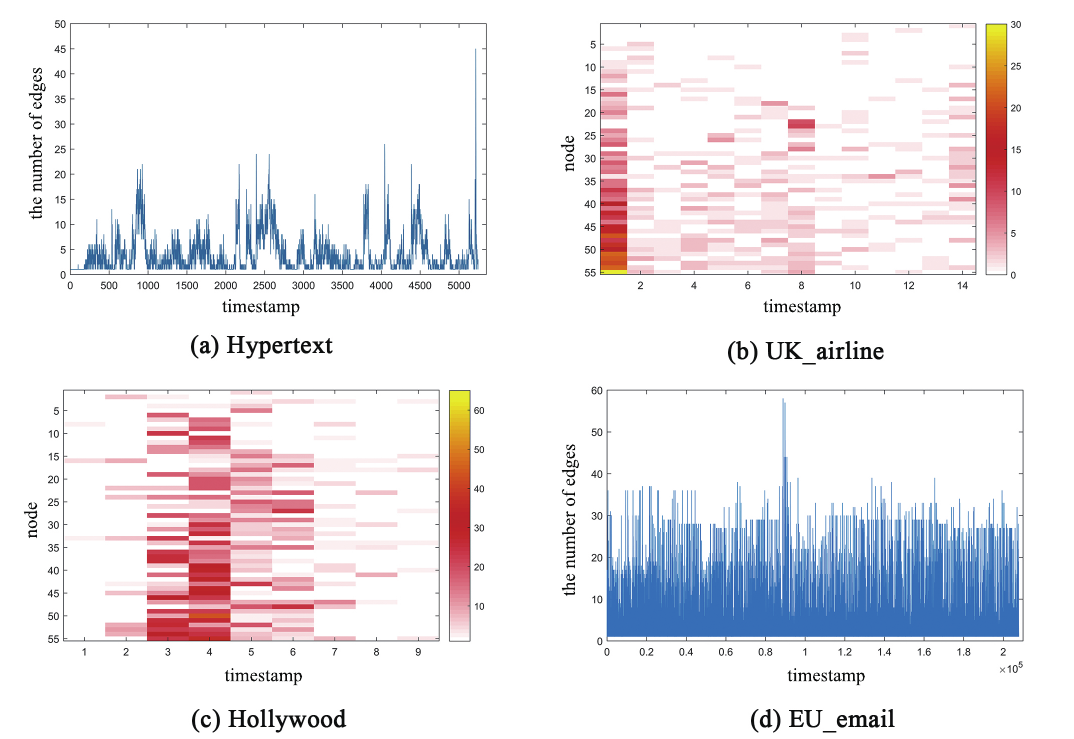}  
\caption{Time vector distribution of four data sets: Due to a large number of timestamps in (a) Hypertext and (d) EU email, the abscissa of the time vector distribution of these two data sets is the timestamp serial number, and the ordinate is the number of interactions. The distribution characteristics of the time vector of the UK Airline (b) and Hollywood (c) data sets are the timestamp serial number on the abscissa and the node serial number on the ordinate (the serial numbers are ranked according to the degree value of the node)}
\label{fig:time vector}
\end{figure*}

To understand the distribution characteristics of the agent behavior at different points in time, we also draw the distribution characteristics of node vectors (shown in Fig.~\ref{fig:time vector}). The agent behavioral characteristics of Hypertext and EU email both show periodicity, which is relevant to the law of human activity (for example, people tend to socialize during the day rather than at night). UK airline, except that agent behavior, is more frequent at time 1 and then more even, whereas the Hollywood actor behavior is concentrated in the middle, which also correlates with the time of an actor's career.

\section{Experiments}
\label{sec:expe}

\subsection{Experimental setup}
\label{sec: experimental setup}
We proposed a new link prediction model with time information named LSTV link prediction model. Therefore, the accuracy of this algorithm should be compared with the LS link prediction model without time information and the LSTD link prediction model with a time decay function. In addition, we proposed the LSTDV model and LSTHV model to distinguish the importance of the time layer from different perspectives and compare them with LSTV model. In the contrast experiments, $\alpha$ was selected to 0.5.

To understand the contribution allocation of timescale similarity and local structure similarity to the link prediction model in different types of networks, we conducted this experiment to observe the potential linking mechanism of different types of networks by adjusting different $\alpha$ values.

\subsection{Experimental datasets}
\label{sec: experimental datasets}
Experiments are conducted on four different datasets: 1) Hypertext 2009~\cite{isella_whats_2011}, 2) UK airline~\cite{morer}, 3) Hollywood~\cite{taylor}, and 4) EU email~\cite{paranjape2017motifs}. To verify the universality of the algorithm, we selected four different timing networks. They differ in network size, timestamp size, time span, and domain. The details are shown in Table~\ref{table:datasets}. Hypertext describes a network of face-to-face contacts of the attendees of the ACM Hypertext 2009 conference. UK airline describes a temporal network of domestic flights operated in the United Kingdom between 1990 and 2003. Hollywood describes a network of all collaborations among 55 actors who were prominent during the Golden Age of Hollywood (1930--1959). EU email is a temporal network that was generated using email data from a large European research institution.
\begin{table*}[htbp]
	\centering
	\caption{\centering \label{table:datasets} Four data sets' attributes and what they represent}
	\begin{tabular}{cccccc} %居中：p{5cm}<{\centering}
		\toprule  % 顶部线
		Dataset & No.\ nodes & No.\ links & No.\ contacts & Time span & Network domain \\ 
		\colrule  % 中部线
		Hypertext & 113 & 2196 & 5246 & 2.5 days & Human proximity network\\
            UK airline & 47 & 406 & 14 & 1990-2003 & Transportation network \\
            Hollywood & 55 & 1043 & 10 & 1909-2009 & Collaboration network \\
            EU email & 986 & 24929 & 206313 & 803 days & Human communication network \\
		\botrule  % 底部线
	\end{tabular}
\end{table*}

\subsection{Experimental results}
\label{sec: experimental results}
The experimental results consist of three parts, which are 1) the comparative experiment between the LSTV link prediction model with time information and the LS link prediction model in static networks, 2) the comparative experiment between the LSTV link prediction model with time vector, the LSTD link prediction model with time decay function, and LSTDV link prediction model with time decay function, and 3) the proportion of the distribution of time and local structure in the link prediction model.

\subsubsection{Comparing the LS and LSTV models}

\begin{figure}
  \includegraphics[width=\columnwidth]{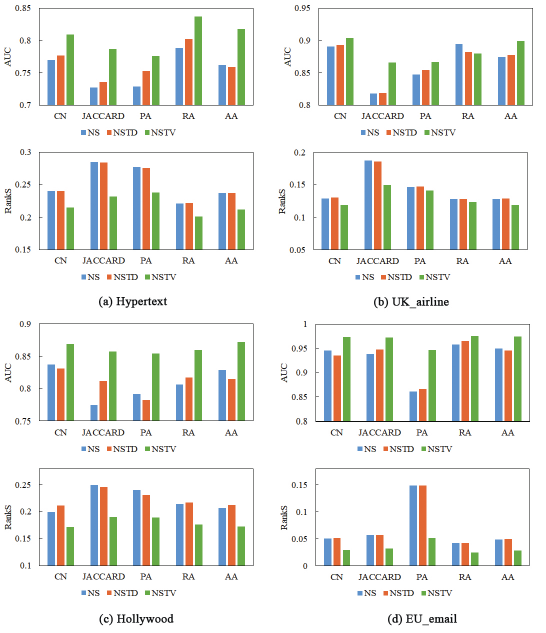}  
\caption{Comparison of the LS, LSTV, and LSTD algorithms by RankS and AUC.}
\label{fig:LS/LSTV/LSTD}
\end{figure}

To explore whether the introduction of time information can supplement the existing link prediction model, we compared the LSTV prediction model with the LS prediction model and LSTD prediction model to observe whether the algorithm accuracy is improved. The experimental results are shown in Fig.~\ref{fig:LS/LSTV/LSTD}.

A low RankS value and high AUC value indicate the high accuracy of the algorithm. It can be seen from the figure that: firstly, compared with LS model and LSTD model, the LSTV algorithm proposed in this chapter shows a better prediction effect in all four data sets. Therefore, we have evidence to believe that the temporal features of the network play an important role in identifying the mechanism of network generation. Secondly, LSTV algorithm maintains the accuracy of different local structure algorithms at a relatively stable and high level. For example, in the dataset of EU email, the algorithm RankS score obtained by using PA index of LS model and LSTD model is very low, which is far lower than the average level of other algorithms. However, LSTV link prediction model brings the accuracy back to the average level. Thirdly, the improvement of LSTV algorithm accuracy is slightly different in different types of data sets. In the human contact network, the accuracy improvement is obvious, while that in the transportation network is the least. This is because the contact behavior of participants in the contact network changes greatly with time, so time has a great effect on the generation and evolution of such networks. The evolution of the transportation network depends more on the distance between two places, economic forces, national relations, and other practical factors. Once a stable structure is formed, it is difficult to change over time. This indicates that time has different effects on different types of networks, and its realistic factors should be considered more carefully in empirical studies. We will discuss the difference in more detail later. Finally, the accuracy of LSTD model and LS model is similar, and it has slight advantages in some situations. This shows that the time decay effect of the local structure effect of the four types of networks is not obvious.

\begin{figure}
  \includegraphics[width=\columnwidth]{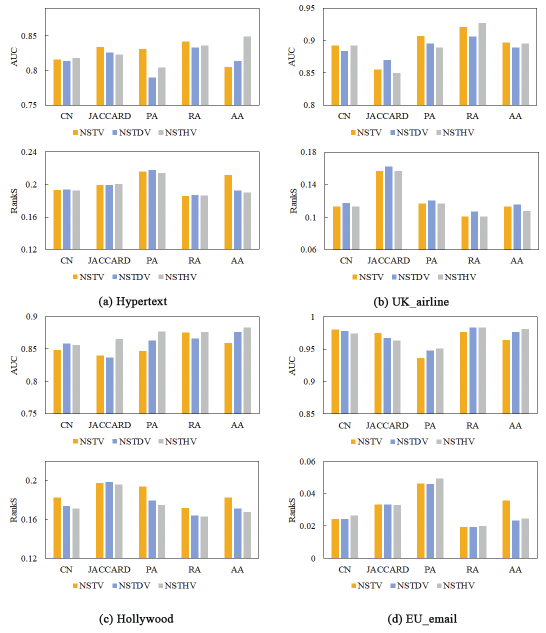}  
\caption{Comparison of AUC and RankS in LSTV, LSTDV, and LSTHV algorithms.}
\label{fig:LSTV/LSTD/LSTHV}
\end{figure}

\subsubsection{Comparing the LSTV, LSTDV, and LSTHV models}

The LSTV model does not consider the importance of different time layers to the topology in computing the timescale similarity. We proposed the LSTDV and LSTHV models that take into account the heterogeneity effect of time layers.  We compared the accuracy of LSTV model, LSTDV model, and LSTHV model, and the experimental results are shown in Fig.~\ref{fig:LSTV/LSTD/LSTHV}.

Compared with the LSTV model, the two models with heterogeneous time effect, LSTDV, and LSTHV showed no significant improvement in most cases. However, in the collaboration network, the accuracy of LSTDV and LSTHV models is improved, especially the prediction model based on CN, PA, and AA indicators. This is because the collaboration network has a large time span and the evolution of the network has a certain dependence on time similarity. From another perspective, the accuracy of LSTHV model is slightly better than that of LSTDV model in most situations, but the gap between the two is not obvious. Therefore, we can ignore the time layer and still achieve a good prediction effect.

\subsubsection{The contribution of timescale similarity and local structure similarity}

In the previous comparative experiments, $\alpha$ was taken as 0.5. In other words, we evenly distributed the contribution of the node's timescale similarity and local structure similarity to the link prediction model. However, we have found in previous experiments that LSTV algorithm improves to different degrees in different types of networks. Therefore, the contributions of time and local structure in the prediction model are different for different types of networks due to different potential linking mechanisms. So, big time or big local structure in the four types of networks?

\begin{figure}
  \includegraphics[width=\columnwidth]{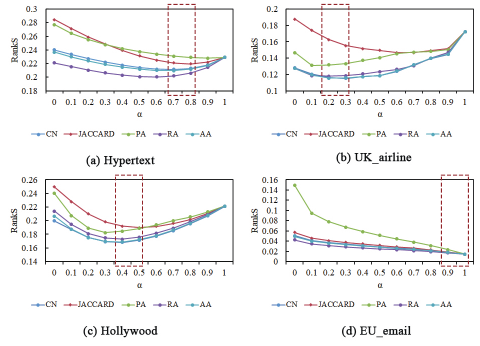}  
\caption{The contribution of timescale similarity and local structure similarity in four data sets.}
\label{fig:contribution}
\end{figure}

Based on these problems, we observe the contribution distribution of time and local structure in the four types of networks by changing the value of $\alpha$. We set the initial value of $\alpha$ to 0 (contribution is devoted all by local structure), the step size is 0.1, and the maximum value is 1 (contribution is devoted all by time). Because AUC has certain random selectivity in the calculation process, the random error will affect the judgment of the evolution trend of the network. Therefore, we selected the RankS index to observe the accuracy of the prediction model. Although the randomly selected probe sets in the prediction model also produce random errors, it is reliable that we choose the same probe sets when adjusting $\alpha$ values. The experimental results are shown in Fig.~\ref{fig:contribution}.

To summarize our experimental results:
\begin{itemize}
\item In the Hollywood collaborative network, when the weight of timescale similarity is about 0.4 and the weight of local structure similarity is about 0.6, the algorithm has the highest accuracy. It shows that in the cooperation network, having a common partner plays a more important role in an agent's choice of cooperation agents. The similarity of the choice of cooperation time also has a greater impact.
\item
In the UK airline aviation network, the accuracy is highest when the contribution of timescale similarity is only 0.2--0.3. This shows that the aviation network is not susceptible to the influence of time and is more dependent on its structural characteristics. This is consistent with the phenomenon that the algorithm accuracy of the temporal link prediction model in the aviation network was not significantly improved in the first experiment.
\item
The situation of Hypertext and EU email is similar. The accuracy trend of the prediction algorithm is almost the same. When the timescale similarity weight is 0.8--0.9, the accuracy is the highest. Moreover, the timescale similarity weight is 1 in the EU email dataset. This indicates that both the timescale similarity of face-to-face communication between people in an unfamiliar environment and the timescale similarity of emails delivered by colleagues between departments affect the choice of the next communication partner. Even in email networks, the effects of local network structures, such as common neighbors and preferred connections, can be ignored. The result also proves that the human proximity network and the human communication network have similar evolutionary characteristics, and empirical research can consider learning from each other's evolutionary models.
\end{itemize}

We found out more about the LSTV model. The RankS value of the model drops faster when $\alpha$ is moved from 0 to 0.1 than after. This once again proves that the addition of timescale similarity can better predict potential links. Among the four networks, the best method of LSTV algorithm combined with local structure similarity is different. This implies that the dynamic mechanism of different types of networks is not completely consistent. In empirical research, we need to design a more suitable prediction model by exploring the evolution mechanism of different networks.

\section{Conclusion}
\label{sec: conclusion}
We proposed the LSTV link prediction model by computing nodes' behavior synchrony as a similarity measure by exploring the temporal characteristics of the network and adding this index to the local structure similarity-based prediction models. Experimental results showed that the timescale similarity index based on nodes' time activity vector could be used as the potential linking mechanism of network generation to predict future links. In the four data sets, the LSTD model does not perform well, which proves that not all networks have a decay effect of local structure similarity. LSTDV and LSTHV are not significantly better than LSTV model: the prediction is not much improved after the heterogeneous time layer effect is compensated for. Therefore, we can ignore the heterogeneous effect of time layers to some extent, whether it is the contribution of time decay or reciprocal link number.

The innovation of this chapter mainly shows as follows: First, we quantified the agents' behavior synchrony index by mining the network time information. Secondly, we proposed an effective and universal temporal link prediction model. This LSTV model not only reflects the full timescale feature of the network but also mines the temporal characteristics of the network. Furthermore, we distinguished the potential linking mechanism of different types of networks through the ratio of different similarity indexes, which is very beneficial to further understanding networks' evolution characteristics. Finally, most previous empirical studies on temporal link prediction have mainly focused on the networks related to human behavior. This chapter is the first to include the evolutionary time characteristics of infrastructure networks in the link prediction model. The experimental results showed that the time characteristic has some weak effect on this kind of network.

Based on the research and methods in this chapter, we believe that time, as a representation of agents' behavior synchrony, can represent a brand-new characteristic of network evolution, and the evolution mechanism and future trend of the entire network can be obtained through the changes of agents' behavior choices. 

The LSTV model shows that the contribution of behaviors is not simply decaying in the full timescale, and some behaviors in the past are still significant. Moreover, our experiment proved that the accuracy of the two models with heterogeneous time effects did not improve significantly. A universal temporal link prediction method, to some extent, can ignore time-layer heterogeneity. However, the prediction results based on different network characteristics and time heterogeneity are more representative of the real world. In empirical studies, how to identify critical time points, assign weights to agents' behavioral synchrony, and the importance of heterogeneous time will be the focus of our future attention.

\begin{acknowledgments}
This research is supported by grants from the National Natural Science Foundation of China (Grant No.\ 42001236, 71991481, 71991480), Beijing Outstanding Talent Training Foundation (Grant No.\ 2018000020124G151). P.H. was supported by JSPS KAKENHI Grant Number JP 21H04595.
\end{acknowledgments}

\bibliographystyle{abbrv}
\bibliography{bib}

\end{document}